\documentclass[a4paper,twocolumn,11pt,accepted=2025-12-30]{quantumarticle}
\pdfoutput=1
\usepackage[utf8]{inputenc}
\usepackage[english]{babel}
\usepackage[T1]{fontenc}
\usepackage[numbers]{natbib}
\usepackage{listings}
\usepackage{amsmath}
\usepackage{amssymb}
\usepackage{amsthm}
\usepackage{hyperref}
\usepackage[resetlabels]{multibib}

\usepackage{algorithm}
\usepackage{algpseudocode}
\usepackage{float}
\usepackage{braket}
\usepackage[shortlabels]{enumitem}
\usepackage{tabularx,booktabs}
\usepackage{comment}

\DeclareMathOperator\atanh{atanh}
\algrenewcommand\algorithmicrequire{\textbf{Input:}}
\algrenewcommand\algorithmicensure{\textbf{Output:}}
\newtheorem{proposition}{Proposition}

\newtheoremstyle{example}
    {3pt}
    {3pt}
    {}
    {}
    {\itshape}
    {:}
    {.5em}
    {}
\theoremstyle{example}
\newtheorem{example}{Example}

\begin{document}
\title{Informed Dynamic Scheduling for QLDPC Codes}

\author{Tzu-Hsuan Huang}
\affiliation{Department of Electrical Engineering, National Tsing Hua University, Hsinchu, Taiwan}
\orcid{0000-0003-3549-0073}
\email{tzuhsuanhuang@gapp.nthu.edu.tw}
\author{Yeong-Luh Ueng}
\email{ylueng@ee.nthu.edu.tw}
\orcid{0000-0002-3438-0385}
\affiliation{Department of Electrical Engineering \& Institute of Communications Engineering, National Tsing Hua University, Hsinchu, Taiwan}

\maketitle

\begin{abstract}
Recent research has shown that syndrome-based belief propagation using layered scheduling (sLBP) can not only accelerate the convergence rate but also improve the error rate performance by breaking the quantum trapping sets for quantum low-density parity-check (QLDPC) codes, showcasing a result distinct from classical error correction codes. 
  
In this paper, we consider edge-wise informed dynamic scheduling (IDS) for QLDPC codes based on syndrome-based residual belief propagation (sRBP). However, the construction of QLDPC codes and the identical prior intrinsic information assignment will result in an equal residual in many edges, causing a performance limitation for sRBP. Two heuristic strategies,  including edge pool design and error pre-correction, are introduced to tackle this obstacle and quantum trapping sets. Then, a novel sRBP equipped with a predict-and-reduce-error mechanism (PRE-sRBP) is proposed, which can provide over one order of performance gain on the considered bicycle codes and symmetric hypergraph (HP) code under similar iterations compared to sLBP.
  
\end{abstract}

\section{Introduction}
In the field of quantum storage, quantum error correction codes are applied to protect the original information from noise by encoding logical quantum bits (qubits) to larger redundant physical qubits. Recently, quantum low-density parity-check (QLDPC) codes, a type of quantum error correction codes, have been a potential candidate for near-term quantum computers for their higher coding rate and minimum distance compared to surface codes \cite{QLDPC,ENCRATE,IBM}. For QLDPC codes, the binary \cite{QuantumBook} or the 4-ary \cite{GF4} isomorphism to elements in Pauli groups allows the application of binary and non-binary syndrome-based belief propagation (sBP) decoders \cite{SparseGraph,IterDec}. The binary sBP is considered in this paper owing to its lower complexity and potential for low-latency hardware implementation.

In order to simultaneously detect both bit-flip and phase-flip errors in quantum error correction, QLDPC codes need to be designed to satisfy orthogonal measurement requirements. However, such construction constraints of quantum codes can result in quantum trapping sets, which will limit decoding performance \cite{QTRAPPING}. For iterative decoders such as sBP, a quantum trapping set refers to a set of variable nodes that that causes the decoder to either fail to converge or converge to an incorrect solution with a syndrome inconsistent with the input, resulting in an error floor. 

In \cite{GBYC,QOSD}, it was shown that by appending the ordered statistics decoder (OSD) to sBP, the error floor can be reduced, indicating a mitigation of the impact of trapping sets. However, the significant error-rate improvement using sBP+OSD comes at the cost of high complexity due to the Gaussian elimination required to solve linear equations, which makes the design of low-latency hardware decoders challenging \cite{FPGA}. The study in \cite{QTRAPPING} demonstrates that using sequential scheduling methods, such as layered sBP (sLBP), can not only accelerate convergence but also reduce the impact of the trapping set, which inspires us to explore different scheduling strategies for sBP based on QLDPC codes.

In classical error correction, unlike fixed sequential scheduling, where the update order is predetermined before the decoder begins, informed dynamic scheduling (IDS) \cite{IDS} is an edge-wise decoder, which can mitigate the impact of classical trapping sets by dynamically changing the update order of edges, further accelerating the convergence. 

A common IDS instance for classical codes is residual belief propagation (RBP) \cite{RBP1}. Each time, RBP updates an edge of the check-to-variable (C2V) node message corresponding to the largest difference between the pre-computed C2V message and the current message, known as the residual. Although RBP can accelerate convergence, its error-rate performance after a sufficiently large number of iterations can be inferior to that of LBP due to the so-called greedy group phenomenon \cite{RBP2}. Specifically, RBP tends to repeatedly update a small subset of edges for certain error patterns, thereby neglecting useful information from other edges, which may lead to convergence failure and performance degradation. Several RBP-based variants have been proposed that alleviate the greedy update of RBP and can further improve the error-rate performance \cite{NWRBP,LMDRBP}.

Although IDS and its variants provide faster convergence for classical LDPC codes, their effectiveness for quantum codes has not been thoroughly studied. Unlike classical error correction, the transmitted information cannot be measured directly, as any measurement would collapse the qubit state \cite{QuantumBook}. Thus, a constant value determined by the fixed channel probability is identically assigned to each variable node as prior intrinsic information. In addition, due to the orthogonal measurement requirements, quantum codes will exhibit more symmetry structures than classical codes. As a result, many edges will have equal residuals, which reduces the chance that sRBP converges to the true syndrome.

To address these challenges for QLDPC codes, we propose two heuristic strategies: 1) Design an edge pool providing diverse edges to enhance the chance of updating edges outside the greedy group or trapping set. 2) Predict and reduce an error position by using the given syndrome information. Based on these strategies, we propose a new sRBP equipped with a predict-and-reduce-error mechanism (PRE-sRBP). In PRE-sRBP, the edge-pool design provides abundant different candidate edges at each update step, while the estimated support sequence further mitigates the impact of trapping sets. Although PRE-sRBP is a heuristic decoding strategy without formal convergence guarantees, it demonstrates promising performance in simulations. For bicycle codes, the first QLDPC codes presented in \cite{SparseGraph}, an elaborately designed edge pool for an sRBP-based decoder enables the sRBP-based decoder to achieve both improved error-rate performance and faster convergence compared to sLBP. For codes containing a large number of quantum trapping sets, such as hypergraph-product (HP) code \cite{HGP}, a more aggressive PRE mechanism is employed to further enhance decoding performance. As a result, the proposed PRE-sRBP can achieve a superior error-rate performance with comparable complexity among the considered sRBP-based decoders across different QLDPC codes. For example, PRE-sRBP can provide an improvement of more than one order of magnitude compared to other sRBP-based decoders on a $[[400,16,6]]$ HP code.

The remainder of this paper is organized as follows. In Section \ref{PRE}, we introduce quantum error correction codes and the sBP. The impact of cycles and quantum trapping sets are also introduced. In Section \ref{SBIDS}, we present several syndrome-based IDS algorithms based on sRBP using the concept of the edge pool. The challenges of using sRBP on QLDPC codes are discussed in Section \ref{issue}, including the details of the greedy group.  In Section \ref{CANDI}, based on these design strategies, the proposed PRE-sRBP is introduced. Section \ref{EXP} shows the simulation results and a complexity analysis. Finally, Section \ref{CONCLU} provides conclusions and outlines future work.

\section{Background}\label{PRE}
\subsection{Quantum Error Codes}\label{PRE:quantum}
In a single-qubit system, Pauli operators describe the operations acting on a single qubit, consisting of $I, X, Z,$ and $Y=iXZ$, where $i=\sqrt{-1}$ and the commutativity is defined by the relations $X^2=Y^2=Z^2=I$ and $XZ=-ZX$ \cite{QuantumBook}. For example, $X$ and $Z$ denote a bit-flip and phase-flip on a single qubit state. The Pauli group $\mathcal{G}_{1}$ is formed by Pauli operators with the multiplicative factors $\pm{1}$ and $\pm{i}$. For an $N$-qubit system, the general Pauli group $\mathcal{G}_{N}$ is considered as the $N$-fold tensor product of $\mathcal{G}_{1}$.

Quantum stabilizer codes are well-studied due to their similar form to classical linear codes \cite{Stabilizer,STAB}. An $[[N,K,d]]$ stabilizer code $C_S$ encodes $K$ logical qubits to an $N$-qubit codeword with a minimum distance $d$. As its name suggests, $C_S$ is defined by its stabilizer group $\mathcal{S}\subseteq \mathcal{G}_{n}$, where $-I\notin\mathcal{S}$. All  elements of $\mathcal{S}$ mutually commute, ensuring that bit-flip and phase-flip errors can be detected simultaneously. Furthermore, $\mathcal{S}$ can be characterized by $N-K$ generators $g_{i}\in\mathcal{G}_{N}, 0\leq i < N-K$,  i.e., $\mathcal{S}=\langle g_{i} \rangle$. An $N$-qubit codeword $\ket{\psi}\in C_S$ is then the $+1$ eigenstate of all the generators, i.e., $\mathcal{P}\ket{\psi}=\ket{\psi}, \ \mathcal{P}\in\langle g_{i} \rangle$. Since the Pauli operators can be expressed as a binary vector, i.e., $I\mapsto (0 \ 0), X\mapsto (1 \ 0), Z\mapsto (0 \ 1), Y\mapsto (1 \ 1)$, the set of $N-K$ generators can then be represented as a binary matrix $H_S=(H_{X_\mathcal{S}} \ H_{Z_\mathcal{S}})$, where $H_{X_\mathcal{S}}, H_{Z_\mathcal{S}}\in F^{(N-K)\times N}_2$. By using the binary representation, the commutativity of each generator is then expressed as the orthogonality of the rows under the symplectic product, i.e., 
\begin{equation}\label{eq:symplectic}
    H_{X_\mathcal{S}}H_{Z_\mathcal{S}}^T\oplus H_{Z_\mathcal{S}}H_{X_\mathcal{S}}^T=\mathbf{0},
\end{equation}
where $\oplus$ is the addition modulo 2. 

It can be observed that identifying two matrices satisfying the constraint (\ref{eq:symplectic}) is not trivial. Thus, a well-known subclass of stabilizer codes, called the Calderbank, Shor-Steane (CSS) codes, are widely used \cite{CSS1,CSS2}. The parity-check matrix for a CSS code has the form 
\begin{equation}
\label{eq:css}
H_{css} = 
 \begin{pmatrix}
H_X & 0 \\
0 & H_Z 
\end{pmatrix}.
\end{equation}
Under this structure, the equation (\ref{eq:symplectic}) can be reduced to $H_XH_Z^T=\mathbf{0}$. Suppose two classical linear block codes satisfy the relationship $C_X\subset C_Z$, then the parity-check matrix $H_Z$ of the code $C_Z$ and $H_X$ of $C^{\bot}_X$ can form a CSS code since $G_XH_Z^T=\mathbf{0}$, where $G_X$ is the generator matrix of $C_X$. If $C_X$ and $C_Z$ are LDPC codes, i.e., the non-zero elements in its parity-check matrix are bound by a constant, then the corresponding quantum code is called a QLDPC code.
The QLDPC codes considered in this paper, including the bicycle code \cite{SparseGraph}, HP code \cite{HGP},  generalized bicycle code \cite{GBYC}, generalized HP code \cite{GBYC}, and bivariate bicycle code \cite{IBM}, all belong to the class of CSS codes.

\subsection{Quantum Channel and Syndrome-based BP (sBP)}\label{secBP}
For an $N$-qubit system, any arbitrary error pattern can be digitalized and represented as an element of the $N$-fold Pauli group, which consists of $X$ and $Z$ \cite{DISCRETE}. Thus, the noisy model can be viewed as two independent channels, a bit-flip with crossover probability $p_x$ and a phase-flip with crossover probability $p_z$. Since all codes considered in this paper are CSS codes and the correlation between $X$ and $Z$ errors is neglected, it is sufficient to decode using just one type of parity-check matrix and error channel. For instance, the bit-flip noisy channel with crossover probability $p_x$ and $H_Z=H$ are considered in this paper.

To apply binary BP on the given error correction code, the parity-check matrix $H\in F^{M\times N}_2$ of the code is first transformed to a factor graph (or the Tanner graph \cite{BP}), which consists of $M$ check nodes and $N$ variable nodes, denoted as $c_i$ and $v_j$, corresponding to the $i$th row and the $j$th column of the parity-check matrix, respectively. An edge is present between check node $c_i$ and variable node $v_j$ if and only if $H_{ij}=1$. Each check node $c_i$ has $d_{c_i}$ edges linked to its neighboring variable node $v_j$, where $v_j\in\mathcal{N}(c_i)=\{v_{j'}|{H}_{ij'}=1\}$ and $d_{c_i}$ is the degree of check node $c_i$. Similarly, each variable node $v_j$ has $d_{v_j}$ edges linked to its neighboring check node $c_i$, where $d_{v_j}$ is the degree of variable node $v_j$ and $c_i\in\mathcal{N}(v_j)=\{c_{i'}|{H}_{i'j}=1\}$. 

To avoid ambiguity in this paper, we refer to an error vector $\mathbf{e}\in F^{N}_2$ as an error pattern. The index of each coordinate of $\mathbf{e}$ is referred to as the error position, or equivalently, error or a variable node (in the Tanner graph). The support of $\mathbf{e}$, denoted as $\text{supp}(\mathbf{e})$, is the set of indices corresponding to the non-zero error positions in $\mathbf{e}$, i.e., $\text{supp}(\mathbf{e})=\{j |0\leq j<N, e_j\neq0\}$. The weight of $\mathbf{e}$ is then defined as the cardinality of $\text{supp}(\mathbf{e})$, denoted as $w(\mathbf{e})$.
The syndrome $\mathbf{s}\in F^{M}_2$ corresponding to an error pattern $\mathbf{e}$ can be computed as $\mathbf{s}=H\cdot\mathbf{e}$. Analogously, the syndrome position or a check node, the support of a syndrome $\text{supp}(\mathbf{s})$, and the weight $w(\mathbf{s})$ are used under a similar definition as for an error pattern.

A key difference between using BP to quantum codes and to classical codes is that the intrinsic information cannot be directly measured, as any measurement would collapse the quantum state. Instead, the syndrome measured on the auxiliary qubits is the only information that can be obtained \cite{QuantumBook}. Consequently, the so-called syndrome-based BP (sBP) is used quantum error correction\cite{SBP}. Considering the sBP decoder in the log domain \cite{LOG}, where the log-likelihood ratio (LLR) is used for message passing, we denote a message from $v_j$ to the neighboring $c_i$ at the $k$th iteration as $L^k_{v_j\rightarrow c_i}$, where $0\leq j<N$.  Then, due to the absence of different prior intrinsic information for each variable node $v_j$, at the $0$th iteration, the prior LLR $L^0_{v_j}$ and $L^0_{v_j\rightarrow c_i}$ are both assigned to the identical constant value $\ln\frac{1-p}{p}$ for all $j$, given $p_x=p$. Then the message from $c_i$ to $v_j$ is computed according to
\begin{align}\label{eq_m_c2v}
    m^{k}_{c_i\rightarrow v_j} &= (-1)^{s_i}\cdot 2\cdot\\&\nonumber
\atanh{\left(\prod_{v_{j'}\in\mathcal{N}(c_i)\setminus v_j}\tanh\left( \frac{L^{k}_{v_{j'}\rightarrow c_i}}{2}\right)\right)},   
\end{align} where $s_i$ denotes the $i$th syndrome position, where $0\leq i<M$.

In the subsequent iterations, $L_{v_j\rightarrow c_i}$ is computed according to
\begin{equation}\label{eq_m_v2c}
    L^{k+1}_{v_j\rightarrow c_i} = L^{0}_{v_j} + \sum_{c_{i'} \in \mathcal{N}(v_j)\setminus c_i}m^{k}_{c_{i'}\rightarrow v_j}	
\end{equation}
The element-wise LLR $L^{k+1}_{v_j}$ at the $(k+1)$th iteration for each $v_j$ can be computed as
\begin{equation}\label{eq_lm}
    L^{k+1}_{v_j} = L^{0}_{v_j} + \sum_{c_i\in \mathcal{N}(v_j)}m^{k}_{c_i\rightarrow v_j}	
\end{equation}
and the estimated error $\hat{e}^{k}_{v_j}=0$ if $L^{k+1}_{v_j}> 0$; Otherwise, $\hat{e}^{k}_{v_j}=1$. Then the estimated syndrome at the $k$ iteration is  $\hat{\mathbf{s}}^{k}=H\cdot\hat{\mathbf{e}}^{k}$. If $\hat{\mathbf{s}}^{k}$ is consistent with the given
syndrome $\mathbf{s}$, the decoder is declared to have converged. Otherwise, the sBP decoder will repeat (\ref{eq_m_c2v}) and (\ref{eq_m_v2c}) until either the decoder converges or the maximum number of iterations $I_{\max}$ is reached. The sBP introduced above is commonly referred to as flooding scheduling since, for an iteration, all the C2V messages are updated, followed by an update of all the variable-to-check (V2C) messages. 

\subsection{Quantum Trapping Sets}\label{PRE:trapping}
In the Tanner graph, a \textit{cycle} is formed by a set of check and variable nodes that create a closed loop, and its length is determined by the number of edges it contains.  As mentioned in Section \ref{PRE:quantum}, CSS codes are constructed from two classical codes that satisfy the symplectic product relationship. This structure results in more cycles, even including length-4 cycles, compared to the classical code, which in turn induces multiple quantum trapping sets. A quantum trapping set consists of classical-type trapping sets and symmetric stabilizer trapping sets. The classical type includes a set of variable nodes that cannot converge correctly even after sufficiently large iterations \cite[Definition 1]{QTRAPPING}. The subgraph induced by such a trapping set typically contains a small number of odd-degree check nodes, with the most problematic trapping sets exhibiting degree-1 and degree-2 check nodes in the induced sub-graphs \cite{TRAPPING}.

In contrast, the symmetric stabilizer trapping sets refer to a set of variable nodes that converge incorrectly, where the corresponding syndrome is inconsistent with the given syndrome \cite[Definition 4]{QTRAPPING}. Symmetric stabilizer trapping, which is unique to quantum codes, is related to code construction. Certain quantum codes have numerous sets of variable and check nodes, whose induced subgraphs have no odd-degree check nodes and can be partitioned into symmetric disjoint subsets \cite[Definition 5]{QTRAPPING}. In such subgraphs, it is possible for two equal-weight error patterns to correspond to the same syndrome. As a result, the iterative decoder struggles to determine which one is better and often converges to the addition of these two error patterns, leading to a zero syndrome. As noted in \cite{QTRAPPING}, HP codes suffer from an error floor due to the abundance of quantum trapping sets. 
\section{Informed Dynamic Scheduling (IDS) for QLDPC codes}\label{SBIDS}
    
If rows (or columns) of the parity-check matrix are divided into different layers and messages are updated layer by layer, this fixed sequential scheduling is referred to as layered BP (LBP)\cite{LBP,SHUFFLE}. For classical codes, LBP can achieve up to twice the convergence rate while maintaining the same error-rate performance. It was shown in \cite{QTRAPPING} that using syndrome-based LBP (sLBP) not only accelerate convergence but also reduce the error floor in QLDPC codes by breaking certain quantum trapping sets, mainly those of the symmetric-stabilizer type. In classical error correction, dynamic scheduling has been demonstrated to further mitigate the impact of trapping sets, which motivates us to explore the potential of syndrome-based IDS for QLDPC codes.

\subsection{Syndrome-based IDS} 

Unlike fixed scheduling, where the update order of the check or variable node is predetermined sequentially before each iteration, IDS is an edge-wise decoder that dynamically determines the update order of edges according to the residual
\begin{equation}\label{gRBP_eq}
r(m_k) =\left \Vert m^{pre}_k-m_k \right \Vert,    
\end{equation}
where the residual is referred to as the norm of the difference between the pre-computed message $m^{pre}_k$ and the current message $m_k$.
For instance, a specific criterion of the RBP \cite{RBP1} is to identify the update order according to the residual of the C2V message, which is defined as 
\begin{equation}\label{RBP_eq}
    r_{c_i\rightarrow v_j} =\lvert m^{pre}_{c_i\rightarrow v_j}-m_{c_i\rightarrow v_j} \rvert.
\end{equation}
The criterion is based on the idea that a larger residual indicates lower message reliability, necessitating more message updates to achieve convergence. Therefore, prioritizing updates to these edges can accelerate the convergence rate. Additionally, following the edge-wise criterion, the edge-wise update order partially relaxes the constraints imposed by the Tanner graph topology compared to fixed scheduling, which may improve the error-rate performance.

To reduce the complexity of the pre-computation $m^{pre}_{c_i\rightarrow v_j}$, the min-sum algorithm (MSA) \cite{NORMINSUM} can be applied to approximate the C2V update in (\ref{eq_m_c2v}) as 
\begin{eqnarray}\label{min_sum}
    m^{k}_{c_i\rightarrow v_j} =(-1)^{s_i}\prod_{v_{j'}\in\mathcal{N}(c_i)\setminus v_j}\text{sgn}(L^{k}_{v_{j'}\rightarrow c_i}) \cdot\\\nonumber
    \min_{v_{j'}\in\mathcal{N}(c_i)\setminus v_j}|L^{k}_{v_{j'}\rightarrow c_i}|.    
\end{eqnarray}

\subsection{The Edge Pool for sRBP}\label{IDSintro}
This section presents several sRBP-based algorithms for quantum codes, introducing the \textit{edge pool} concept to highlight their similarities and differences. Let $c_{\max}$ and $v_{\max}$ respectively denote the chosen check node and the variable node corresponding to $\max_{\bigl\{\{c_i,v_j\}\in \mathfrak{R}\bigl\}} r_{c_i\rightarrow v_j}$, where $\{c_i,v_j\}$ represents an edge connected by check node $c_i$ and variable node $v_j$, and an edge pool $\mathfrak{R}$ is a set of edges. Then, after computing all residuals $r_{c_i\rightarrow v_j}$ for all $0\leq i<M,  v_j\in\mathcal{N}(c_i)$ by using (\ref{RBP_eq}) and (\ref{min_sum}), we can arrange the sRBP and its variants into three main steps:
\begin{enumerate}[label=\textbf{S.\arabic*}]
    \item\label{s:s1} Select the $c_{\max}$ and $v_{\max}$ corresponding to the maximum residual of all edges in the edge pool $\mathfrak{R}$ and update the edge from $c_{\max}$ to $v_{\max}$ by (\ref{eq_m_c2v}).
    The residual of the updated edge is assigned as zero. 
    
    \item\label{s:s2} For every $c_a\in\mathcal{N}(v_{\max})\setminus c_{\max}$, update edges from $v_{\max}$ to $c_a$ using (\ref{eq_m_v2c}), and for every $v_b\in\mathcal{N}(c_a)\setminus v_{\max}$, compute the new residual of the edges from $c_a$ to $v_b$ using (\ref{RBP_eq}).
    
    \item\label{s:s3} Select the next $c_{\max}$ and $v_{\max}$ as in \ref{s:s1} but from the new edge pool $\mathfrak{R}'$, which depends on different decoders.
\end{enumerate}
The decoder will repeat the three steps above for a single iteration. A single iteration is conventionally defined as the sequence of updates in which the number of C2V updates reaches the total number of undirected edges $E$ in the parity-check matrix due to an unbalanced update of the edge-wise decoder. It can be observed that \ref{s:s2} only depends on the choice of $c_{\max}$ and $v_{\max}$ in \ref{s:s1}. As a result, different IDSs concentrate on designing the $\mathfrak{R}$ and $\mathfrak{R}'$. The design for the sRBP is shown in Algorithm \ref{alg:RBP}. 

\begin{algorithm}[H]
\caption{Edge pool for sRBP}
\label{alg:RBP}
\begin{algorithmic}
\State In \ref{s:s1}, the edge pool consists of edges from all check nodes, i.e.,
\begin{align*}
    \mathfrak{R}=\Bigl\{\{c_i,v_j\}|0\leq i<M,   v_j\in\mathcal{N}(c_i)\Bigl\}
\end{align*}
And the edge $m_{c_{\max}\rightarrow v_{\max}}$ is updated.
\State In \ref{s:s3}, the edge pool $\mathfrak{R}'=\mathfrak{R}$.
\end{algorithmic}
\end{algorithm}

In classical error correction, the convergence rate of RBP is faster than for both BP and LBP, yielding improved error-rate performance during the initial iterations. However, for a large number of iterations, RBP performance may deteriorate due to the \textit{greedy group phenomenon} \cite{RBP1}. Specifically, when RBP encounters an error that it can not solve, it may attempt to update the incorrect variable node, leading to additional unsatisfied check node. As a result, the residuals of edges connected to this incorrect variable node will remain large. This behavior results in excessive computational effort concentrated on a small subset of edges, often preventing successful convergence.

Two algorithms have been proposed to address this issue. A node-wise RBP (NW-RBP) \cite{NWRBP} simultaneously updates all edges emanating from the same check node, allowing the update of edges outside the greedy group and ultimately improving the error-rate performance. Specifically, suppose a variable node within a trapping set has been corrected. In that case, at least one degree-2 check node becomes a degree-1 node, which is likely to be chosen next due to its large residual. This, in turn, increases the probability of correcting another variable node within the trapping set.

Another algorithm is known as the latest-message-driven RBP (LMD-RBP) \cite{LMDRBP}, in which the authors propose a strategy where the next update is related to the latest-updated variable node. Based on the reason that after updating $m_{c_{\max}\rightarrow v_{\max}}$ and propagating the related edges from $v_{\max}$ to its neighboring $c_a$, not only the message for the edge from $v_{\max}$ but also the message emitting from $c_a$ will be more reliable. As shown in\cite{LMDRBP}, LMD-RBP algorithm exhibits better performance during the initial iterations for 5G New Radio LDPC codes \cite{5G}. Syndrome-based NW-RBP (NW-sRBP) and the LMD-RBP (LMD-sRBP), modified and used for quantum codes in this paper, are shown in Algorithm \ref{alg:NWRBP} and Algorithm \ref{alg:LMDRBP}, respectively. 

\begin{algorithm}[H]
\caption{Edge pool for NW-sRBP}
\label{alg:NWRBP}
\begin{algorithmic}[1]
\State In \ref{s:s1}, the edge pool passes through the edges from all check nodes, i.e.,
\begin{align*}
    \mathfrak{R}=\Bigl\{\{c_i,v_j\}|0\leq i<M,   v_j\in\mathcal{N}(c_i)\Bigl\},
\end{align*}
and the messages $m_{c_{\max}\rightarrow v_k}$ are updated for all $v_k\in\mathcal{N}(c_{\max})$.
\State In \ref{s:s3}, the edge pool $\mathfrak{R}'=\mathfrak{R}$.
\Statex {\bf Note:} In \ref{s:s1}, edges from $c_{\max}$ to $v_k\in\mathcal{N}(c_{\max})$ are updated simultaneously, while \ref{s:s2} and \ref{s:s3} remain unchanged, except for replacing $v_{\max}$ with $v_k$.
\end{algorithmic}
\end{algorithm}

\begin{algorithm}[H]
\caption{Edge pool for LMD-sRBP}
\label{alg:LMDRBP}
\begin{algorithmic}[1]
\State In \ref{s:s1}, the edge pool passes through the check nodes, i.e.,
\begin{align*}
    \mathfrak{R}=\Bigl\{\{c_i,v_j\}|0\leq i<M,   v_j\in\mathcal{N}(c_i)\Bigl\},
\end{align*}
and the message $m_{c_{\max}\rightarrow v_{\max}}$ is updated.
\State In \ref{s:s3}, the edge pool $\mathfrak{R}'$ is obtained by the new $v^*_{\max}$ from $\mathfrak{R}_I$, i.e.,
\begin{equation*}
\begin{split}
 \mathfrak{R}_I = \Bigl\{\{c_i,v_j\}|c_i\in\mathcal{N}(v_{\max})\setminus c_{\max} \\
 v_j\in\mathcal{N}(c_i)\setminus v_{\max}\Bigl\}\\
\mathfrak{R'} =\Bigl\{\{c_i,v^*_{\max}\}|c_i\in\mathcal{N}(v^*_{\max})\Bigl\}  
\end{split}
\end{equation*}
\end{algorithmic}
\end{algorithm}

\subsection{Error Weight Analysis}
\begin{figure}[ht]
  \centering
\includegraphics[width=\columnwidth]{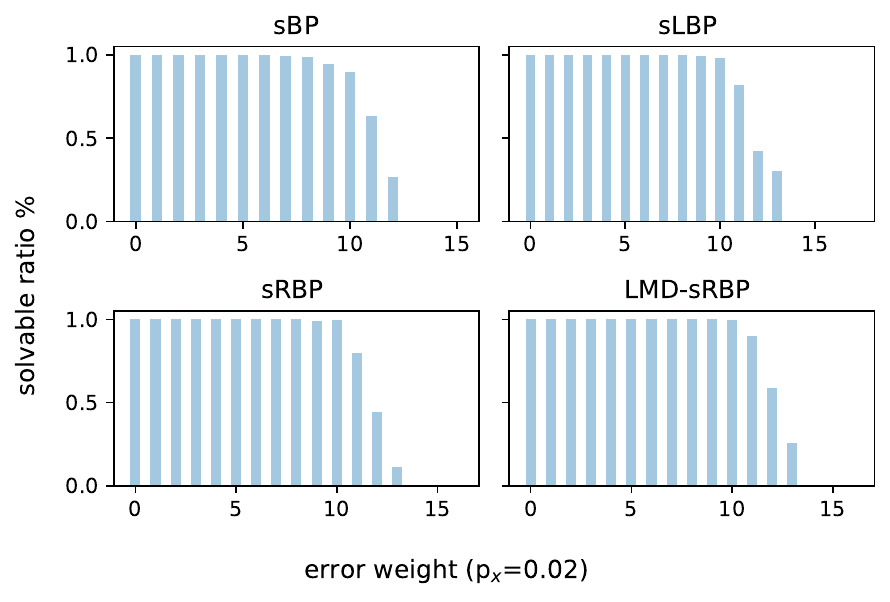}
  \caption{The solvable ratio of different error weights using sBP, sLBP, sRBP, and LMD-sRBP where $I_{\max}=90$.}
  \label{fig:errWeightRBP256}
\end{figure}

Fig.~\ref{fig:errWeightRBP256} shows the distribution of the error weight versus the solvable error ratio using sBP, sLBP, and other sRBP variations on the $[[256,32]]$ bicycle code.
For the analysis, all error patterns are collected until the total number of error frames reaches 100, from which the solvable ratio is calculated. For example, at \( p_x = 0.02 \), there are 117 error patterns with weight 10, of which 105 can be correctly decoded by sBP. Therefore, the solvable error ratio for this weight is $89\%$. 

The results show that sBP can effectively solve error patterns with weight up to 9, achieving a solvable ratio exceeding $94\%$. However, its performance deteriorates for higher weights, and patterns with weight greater than 13 cannot be solved. In comparison, sLBP can decode error patterns with weight up to 10 with a ratio above $98\%$. sRBP further improves the performance, achieving a solvable ratio over $99\%$ for weights up to 10, though the ratio drops rapidly for weights greater than 11. In contrast, by using LMD-sRBP, almost all error patterns with weight 10 or less can be solved with a ratio above $99\%$, and the average solvable ratio for weights above 11 is higher than that of sLBP.

It is also worth noting that since sBP-based decoders are not bounded-distance decoders, their decoding capacity is influenced by the graph structure and bit-flip probability $p_x$. For instance, the $[[256, 32]]$ bicycle code has an unknown distance, but it is less than 16, as the distance of bicycle codes is limited by their row weight \cite{SparseGraph}. Simulation results indicate that the solvable weight using sBP exceeds half of this upper bound distance, although in a probabilistic sense. This can also be observed in the $[[144, 12, 12]]$ bivariate bicycle code, where each component classical code has distance 12 \cite{IBM}. Specifically, simulations (not shown) indicate that $92\%$ of error patterns with weights up to 10 can be decoded successfully using sBP at \( p_x = 0.02 \), with a maximum number of iterations \( I_{\max} = 90 \).

\section{Challenges for sRBP on QLDPC codes} \label{issue}
\subsection{Greedy Group Effect on sRBP} \label{GREEDY_EX}

\begin{figure}[ht]
	\includegraphics[page=2,width=1.2\columnwidth]{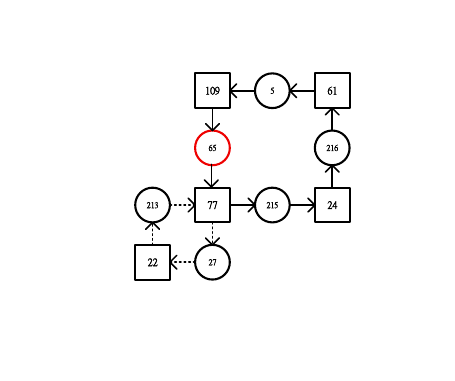}
	\caption{Greedy update example. Circles and squares represent variables and check nodes, respectively.}
	\label{GREEDY}
\end{figure}

Section \ref{IDSintro} describes the greedy group phenomenon when using RBP on classical codes, which is also observed for sRBP on quantum codes. For instance, consider the \([[256,32]]\) bicycle code, a type of QLDPC code. Given the syndrome corresponding to the error pattern with support $\{46, 63, 93, 115, 122, 161, 214, 229, 232, 240, 244,\\ 253\}$, sRBP tends to misidentify variable node $65$ due to the prioritization based on its syndrome support.
After 16 iterations, a loop involving variable node $65$ emerges, consuming a substantial number of update resources. Figure~\ref{GREEDY} illustrates this loop, where circles and squares represent the variable and check nodes, respectively. It should be noted that the loop does not imply a specific update order; for instance, after updating the edge from $109$ to $65$, the edge from $77$ to $27$ might be updated several rounds later. While the loop involving node $65$ may not be unique, the number of updates to incorrect variable nodes within such loops is considerably higher than that for the actual variable node in error.

In the original sRBP, the edge pool $\mathfrak{R}$ encompasses all edges of the entire graph of the code. Consequently, a selected edge may appear in subsequent selection rounds and be chosen multiple times. If the selected edge can reliably transmit a message to an unreliable variable node, the convergence rate will be improved. However, if the edge leads to an incorrect variable node, such as variable node $65$ in Fig. \ref{GREEDY}, repeated updates of the same edges consume significant computational resources without achieving convergence.

This issue can be addressed by using diverse edge pools. By varying the composition of the edge pool in each selection round, the accumulation of updates within loops or cycles may be altered. In the previous example, the incorrect estimate at variable node $65$ may be corrected after several iterations when using NW-sRBP or LMD-sRBP. However, both decoders still fail to converge for this syndrome. Therefore, new designs need to be explored for QLDPC codes to further enhance the decoding performance.

\subsection{Quantum Error Correction Constraint} \label{DESIGN}
As discussed in Section \ref{PRE:trapping}, the presence of trapping sets in quantum codes adversely affects the performance of the original sBP. Although using IDS instead can mitigate the impact of trapping sets on classical codes, its effectiveness on quantum codes is limited due to the absence of distinct prior information. In classical codes, the distinct prior information provide biased and accurate updates on edges from the first iteration. For example, the residual of variable node $r(v_j)$, computed from its distinct prior LLR $L^{0}_{v_j}$, can be used to assist the decoder in identifying the reliability of edge messages and adjust the update  priority accordingly \cite{DYNAMIC_SELECTION, TABU, RELIABLE_METRIC}. However, this assistance design is limited when considering syndrome-based IDS on QLDPC codes since the identical prior LLR is assigned for all variable nodes as described in Section \ref{secBP}. 

Furthermore, QLDPC codes are often regular, defined in \cite{CLDPC} as the parity-check matrix having a constant row weight and a constant column weight. Here, weight refers to the number of non-zero positions in a row or column. This regularity of QLDPC codes, due to construction constraints introduced in Section \ref{PRE:quantum}, coupled with identical prior LLR assignments, results in an equal residual in many edges. Thus, the decoder may select edges that do not contribute to convergence, wasting computational resources. Worse, if the incorrectly selected variable node belongs to a trapping set or cycle, the decoder may fail to address the impact of the trapping set, imposing a performance limitation. For example, as illustrated in Fig. \ref{GREEDY}, if variable node $213$ in the 4-cycle is incorrectly decoded after the edge from $22$ to $213$ has been updated, the edge from $77$ to $27$ might also be updated, since the residuals of these edges remain significant when the involved variable nodes are not actual errors.

Under dynamic scheduling, certain edges may never be selected, implying that they are dismissed from the graph. Without distinct prior information, edge selection can thus lack accuracy. Tracking the number of updates per edge in NW-sRBP or LMD-sRBP reveals that some edges remain unupdated even after 16 iterations, potentially preventing convergence to the given syndrome. Therefore, a key design direction is to involve a broader set of edges in the update process, facilitating the collection of more related internal information while maintaining biased and informative distribution of updates.


\section{The Proposed sRBP for QLDPC codes}\label{CANDI}

To address the negative impact of the greedy group and trapping set, we propose two heuristic strategies. The first strategy is to design an edge pool that can offer various edges whenever a selection is made in \ref{s:s1}. By increasing the variety of edges considered, the chances of information exchange both within and outside the greedy group or trapping set are enhanced, which can improve the probability of successful decoding. 

The second strategy is to predict an error position likely belonging to the support of an error pattern and perform a pre-correction. If an error position within a trapping set is corrected early, the influence of the trapping set may be mitigated. However, a candidate error position should be elaborately found since the syndrome is used as input; otherwise, the syndrome corresponding to the new error pattern after pre-correction may differ significantly from the original, limiting decoder performance.

\subsection{A Proposed Edge Pool $\mathfrak{R}$ for QLDPC Codes}\label{CQRBP}

To provide the diversity of the edge pool, we propose an $\mathfrak{R}$ induced by a variable node $v_t$ and a list of its neighboring check nodes that have not yet been updated, where $v_t$ will be adjusted for each edge selection.

When considering a variable node \( v_t \), only edges connected to \( v_t \) are compared in Step \ref{s:s1} to determine which C2V edge to update. Here, \( v_t \) serves as an indicator and is updated for the subsequent edge selection, ensuring that the edges in \( \mathfrak{R}' \) differ from those in \( \mathfrak{R} \). In simulations, \( v_t \) is assigned sequentially from \( 0 \) to \( N-1 \). 

An additional $d_{v_t}$-tuple flag list $\mathbf{f}^{v_t}$ is further introduced to the edge selection for each $v_t$. Each element in $\mathbf{f}^{v_t}$, denoted as $f^{v_t}_{c_i}$, corresponds to neighboring check node $c_i\in\mathcal{N}(v_t)$, which is initialized to 0. When selecting a C2V edge from the pool induced by $v_t$, only edges connected to check nodes with $f^{v_t}_{c_i}=0$ are considered. 
Once the C2V edge is updated in \ref{s:s1}, we set $f^{v_t}_{c_{\max}}=1$. This mechanism further increases the diversity of edges between $\mathfrak{R}$ and $\mathfrak{R}'$ induced by $v_t$. To ensure at least one edge can be updated from $\mathfrak{R}$ induced by $v_t$, all $f^{v_t}_{c_i}$ are reset to 0 once the number of neighboring check nodes with $f^{v_t}_{c_i}=1$ reaches $d_{v_t}-1$.  

The proposed edge pool refines the global search of the original sRBP into a local search centered on each $v_t$. The flag $\mathbf{f}^{v_t}$ ensures that every edge receives at least one update, as outlined in Section \ref{DESIGN}. Overall, this design not only allows a more comprehensive exploration of the code structure but also preserves a biased update distribution of the edges. 
Using the proposed edge pool, the decoder can efficiently resolve the true error pattern discussed in Section \ref{GREEDY_EX}. The design for $\mathfrak{R}$ and $\mathfrak{R}'$ is summarized in Algorithm \ref{alg:CQRBP}.

An additional advantage is that the complexity of the proposed design can be lower than the sRBP-based decoders presented in Section \ref{SBIDS}, as only the C2V edges neighboring $v_t$ and the check nodes with $f^{v_t}_{c_i}=0$ will be involved and selected after the residual comparison. 

\begin{algorithm}[H]
\caption{Proposed edge pool design}
\label{alg:CQRBP}
\begin{algorithmic}[1]
\State In \ref{s:s1}, the edge pool covers check nodes neighboring $v_t=0$ that have not yet been updated, i.e.,
\begin{align*}
\begin{split}
    \mathfrak{R}=\Bigl\{\{c_i,v_j\}|c_i\in\mathcal{N}(v_t), f^{v_t}_{c_i}=0,\\
    v_j\in\mathcal{N}(c_i)\setminus v_t\Bigl\}
\end{split}
\end{align*}
and the message $m_{c_{\max}\rightarrow v_{\max}}$ is updated. Set $f^t_{c_{\max}}=1$.
\State In \ref{s:s3}, the edge pool is induced by the new $v_l=v_t+1,v_l<N$, i.e., 
\begin{align*}
\begin{split}
    \mathfrak{R}'=\Bigl\{\{c_i,v_j\}|c_i\in\mathcal{N}(v_l), f^{v_l}_{c_i}=0,\\
    v_j\in\mathcal{N}(c_i)\setminus v_l\Bigl\}
\end{split}
\end{align*}
\end{algorithmic}
\end{algorithm}

\subsection{Error Prediction from the Syndrome}\label{ESTIMATED}
The second strategy aims to perform a prediction of the possible error position prior to the executing the proposed sRBP. To predict a candidate error belonging to $\text{supp}(\mathbf{e})$ given the syndrome $\mathbf{s}=H\cdot\mathbf{e}$, Proposition \ref{prop} is provided to assist us in identifying the candidate more accurately.
\begin{proposition}\label{prop}
  Considering a QLDPC code where $H_Z=H$ in which the check node degree is a constant $d_c$ over the bit-flip channel with crossover probability $p_x = p$, a priority of the estimated error position belonging to the $\text{supp}(\mathbf{e})$ given the syndrome $\mathbf{s}$ can be determined by the value $d_{v_j}-2w^1_{v_j}$, where $d_{v_j}$ is the degree of $v_j$, and $w^1_{v_j}$ denotes the cardinality of the set of check nodes neighboring $v_j$ and belonging to the support of the given syndrome, i.e., $w^1_{v_j}=|\{c_i|c_i\in \mathcal{N}(v_j),i\in\text{supp}(\mathbf{s})\}|$.
\end{proposition}\label{statment}
\begin{proof}
    The proof is detailed in Appendix \ref{APPENDIX2}.
\end{proof}

The lower the value of $d_{v_j}-2w^1_{v_j}$ means the higher the ratio of the check nodes neighboring $v_j$ involving the given syndrome, which implies that the $j$th error position is likely belonging to $\text{supp}(\mathbf{e})$. Candidate error positions can thus be obtained by computing this metric for all $v_j$ and sorting in ascending order. A case is presented in Example \ref{ex:errorC}. 

\begin{example}\label{ex:errorC}
Considering the $[[256,32]]$ bicycle codes with the given syndrome $\mathbf{s}$, where  $\text{supp}(\mathbf{s})=\{2,7,20,22,23,26,28,38,40,44,50,54,55,60,75,\\
80,81,88,90,91,94,97\}$. By computing $d_{v_j}-2w^1_{v_j}$ for each $v_j$ and sorting in ascending order of negative numbers, we can obtain the sequence $\{31,43,93,115,151,9,25,33,...\}$, with the corresponding $d_{v_j}-2w^1_{v_j}=\{-5,-4,-3,-2,-1,0,0,0,...\}$. 
The error pattern $\mathbf{e}$ corresponding to $\mathbf{s}$ has $\text{supp}(\mathbf{e})=\{31,43,93,115\}$, which are exactly the first $4$ elements in this sequence.
\qed
\end{example}

Based on Proposition \ref{prop}, an estimated support sequence can be constructed to reflect the possibility of each variable node belonging to $\text{supp}(\mathbf{e})$ given a syndrome, where the most likely error position in $\text{supp}(\mathbf{e})$ will occur at the beginning of this sequence. Since not every element in the support of an error pattern will exist in the first few positions in the estimated support sequence, we must carefully reduce the effect of an error position in the sequence individually. If a selected position indeed belongs to $\text{supp}(\mathbf{e})$, the corresponding syndrome entries can be subtracted from the given syndrome, potentially facilitating successful decoding by the proposed sRBP. Conversely, if the selected position is not part of the true support, the next position in the sequence is considered for a subsequent decoding attempt.

\subsection{The Proposed sRBP with a Predict-and-reduce-error Mechanism (PRE-sRBP)}\label{PROPOSED}

Now, we introduce the proposed algorithm. Denote $\lambda_{\max}$ as the maximum number of candidate error positions that we want to select from the estimated support sequence. Each selection of a candidate error position is referred to as a \textit{trial}. The algorithm begins with the first position in the estimated support sequence, corresponding to $\lambda=0$. 

For each trial we record the $N$-tuple standard error pattern $\mathbf{e}_c=(0,...0,1,0,...0)$, where only the $c$th position is set to $1$ and $c$ is the $\lambda$th element in the estimated support sequence. To reduce the impact of $\mathbf{e}_c$, its corresponding syndrome is computed as $\mathbf{s}_{\mathbf{e}_c}=H\cdot\mathbf{e}_c$ . The input syndrome for the decoder presented in Section \ref{CQRBP} is then obtained by $\mathbf{s}_{\mathbf{r}}=\mathbf{s}\oplus\mathbf{s}_{\mathbf{e}_c}$. The decoder for each trial will operate $I_t$ iterations; thus, the total maximum number of iterations $I_{\max}$ will become $I_t\cdot \lambda_{\max}$.   

If the decoder converges to an estimated error pattern $\hat{\mathbf{e}}_r$, the decoding process is terminated and the final estimated error pattern is given by $\hat{\mathbf{e}}=\hat{\mathbf{e}}_r+\mathbf{e}_c$. If the decoder fails to converge within $I_t$ iterations, the algorithm proceeds to the next position in the estimated support sequence, where $0 \leq\lambda  < \lambda_{\max}$. This trial-based mechanism can also be executed in parallel to reduce decoding latency. If none of the $\lambda_{\max}$ trials converges, the decoder declares a failure. In summary, the proposed sRBP is equipped with a \textit{predict-and-reduce-error} mechanism and is therefore referred to as PRE-sRBP. The complete procedure is summarized in Algorithm \ref{alg:alg5}. The function code can be found on github \cite{SOURCE}.

\subsection{Key Differences from Existing Decoding Algorithms for Classical Codes} 
In this section, the proposed technique used in PRE-sRBP will be compared to existing augmented BP \cite{ABP} and Chase decoder \cite{CHASE} used in classical error correction codes. 

In augmented BP \cite{ABP}, multiple selection rounds occur throughout the iterations. In each round, a specific fixed value is assigned to a variable node. Thus, the choice of the variable node in each round impacts subsequent selections, requiring the node selection algorithm in \cite{ABP} to perform iterative comparisons for accurately identifying the target variable node. In contrast, our PRE mechanism operates in a one-shot manner. Specifically, the PRE mechanism identifies $\lambda$ candidate error positions and eliminates them from each trial before employing the proposed sRBP decoder. In other words, the variable nodes selected in each trial are independent. This leads to a reduction in computational complexity. By executing the PRE mechanism only once and running $\lambda$ trials of the proposed sRBP decoder in parallel, the decoding latency can be further reduced, making the proposed approach particularly attractive for hardware implementation.

We now compare our PRE-sRBP to the Chase decoder used in classical codes \cite{CHASE}.
The Chase decoder evaluates various test error patterns, where the positions of the 1's are determined by the channel LLR. The objective is to adjust the received word so that it is closely aligned with the nearest codeword by adding a small set of those selected error patterns. The key differences between these two decoders are outlined below.
\begin{enumerate}
    \item In the Chase decoder, to reduce the number of test patterns, the confidence value (or LLR) is used to determine the positions of 1's in the test patterns. Since there is no prior information available for quantum codes, the proposed decoder uses the proposition to obtain hard syndrome information to predict a candidate error instead of using soft channel information.
    
    \item 
    In the Chase decoder, the number of test patterns and the locations of 1's are related to the capacity of the bounded-distance decoder, specifically \(\lfloor (d-1)/2 \rfloor\), where $d$ is the minimum distance of the code. However, since the decoding capacity of the proposed sRBP decoder is not only determined by the distance, the parameter \(\lambda_{\max}\) can be configured independently and is not tied to the specific codes used. In addition, since it is the syndrome $\mathbf{s}_r$ as the input for the proposed sRBP, a weight-one standard error pattern $\mathbf{e}_c$ is used for each trial to avoid the large Hamming distance between $\mathbf{s}_r$ and the original syndrome $\mathbf{s}$.  
\end{enumerate}

\begin{algorithm}[H]
\caption{Proposed PRE-sRBP}\label{alg:alg5}
\begin{algorithmic}[1]
\def\NoNumber#1{{\def\alglinenumber##1{}\State #1}\addtocounter{ALG@line}{-1}}
\Require
{$H\in F^{M\times N}_2$, $\mathbf{s}\in F^{M}_2$, bit-flip channel crossover probability $p_x=p$, and the maximum selection number $\lambda_{\max}$.}
\Ensure The estimated error $\hat{\mathbf{e}}$ 
\State Create the estimated support sequence: Compute  $(d_{v_j}-w^1_{v_j})$ for all $v_j$ and sort in ascending order. Set the parameter $\lambda=0$.
\State Initialize $L^{0}_{v_j}=\ln\frac{1-p}{p}$ and $L^{(0)}_{v_j\rightarrow c_i}$ using $L^{0}_{v_j}$ for all $v_j$ and $c_i\in\mathcal{N}(v_j)$. Set the maximum number of iterations $I_t$ for each trial $\lambda$ and the iteration index $k=0$. 
\State Reduce the impact from $\mathbf{e}_c$ by computing $\mathbf{s}_{\mathbf{e}_c} = \mathbf{e}_c\cdot H^T$ and the input syndrome $\mathbf{s}_{\mathbf{r}}=\mathbf{s}\oplus\mathbf{s}_{\mathbf{e}_c}$, where $c$ is the $\lambda$th element in the estimated support sequence. 

\State Apply the design for $\mathfrak{R}$ in Section \ref{CQRBP} using $\mathbf{s}_{\mathbf{r}}$ and $k<I_t$:
\State~~~Compute $r_{c_i\rightarrow v_j}, 0\leq i<M,   v_j\in\mathcal{N}(c_i)$.
\State~~~Initialize the flag list $\mathbf{f}^{v_t}$. Set $v_t=0$.
\State~~~Select the $c_{\max}$ and $v_{\max}$ in $\mathfrak{R}$, where \Statex~~$\mathfrak{R}=\{c_i,v_j|c_i\in\mathcal{N}(v_t), f^{v_t}_{c_i}=0, v_j\in\mathcal{N}(c_i)\setminus v_t\}$ and follow  step \ref{s:s1}.
\State~~~Set $f^{v_t}_{c_{\max}}=1$ and follow step \ref{s:s2}. 
\State~~~Set $v_t=v_t+1, v_t<N$ and follow step 
\Statex~~~\ref{s:s3} by using $\mathfrak{R}=\mathfrak{R}'$.
\State~~~When the updated C2V number reaches $E$:
\State~~~~{\bf If} converge to $\hat{\mathbf{e}}_r$, {\bf return} $\hat{\mathbf{e}}=\hat{\mathbf{e}}_r+\mathbf{e}_c$ 
\State~~~~{\bf Else if} $k<I_t$, {\bf then} $k = k+1$
\Statex~~~~~~~~~~~~~~and go to 7.
\State~~~~{\bf Else} Set $\lambda = \lambda+1,  \lambda<\lambda_{\max}$ and go to 2.
\State {\bf return} failure.
\end{algorithmic}
\end{algorithm} 

\subsection{Error Weight Analysis}\label{PRE_EX}
\begin{figure}[ht]
  \centering
  \includegraphics[width=0.99\columnwidth]{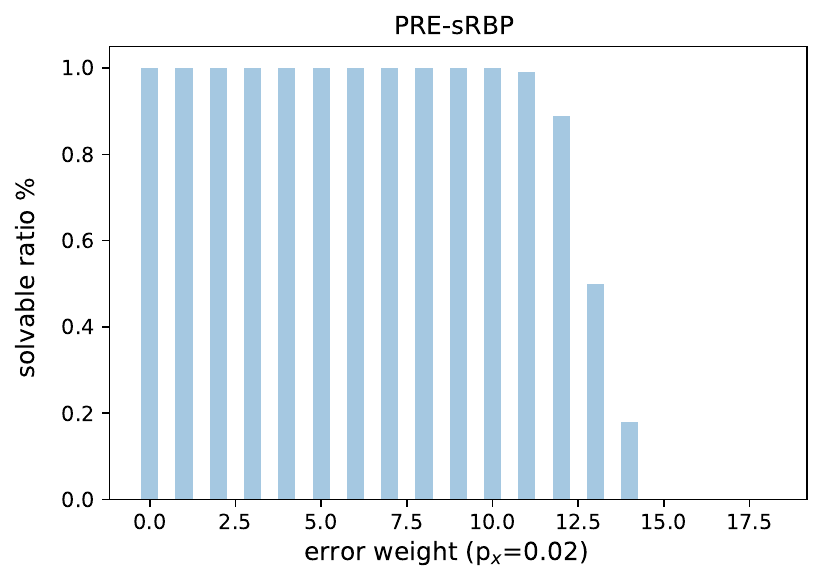}
  \caption{The solvable ratio for different error weights by using PRE-sRBP where $ \lambda_{\max}=15$ and $I_t=6$.}
  \label{fig:errWeightPRO256}
\end{figure}

Fig.~\ref{fig:errWeightPRO256} shows the solvable ratio using PRE-sRBP where $\lambda_{\max}=15$ and $I_t=6$. Compared with the results in Fig.~\ref{fig:errWeightRBP256}, PRE-sRBP can solve error patterns with weights up to $11$. However, the solvable ratio decreases to $88\%, 50\%,$ and $17\%$ for weights of $12, 13$, and $14$, respectively. 

The mitigation of trapping-set effects is possible if a variable node within a trapping set is selected based on the estimated support sequence. A scenario is demonstrated in Example \ref{ex:select} to show the effect of such variable-node selection on the decoder.

\begin{figure}[htb]
  \centering
  \includegraphics[page=1,width=0.99\columnwidth]{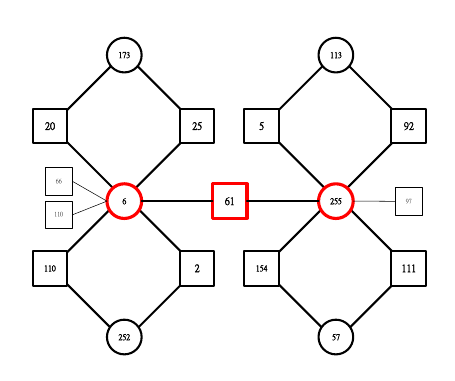}
  \caption{A trapping set example. Those colored in black and red indicate a variable node (check node) that belongs to and does not belong to the support $\text{supp}(\mathbf{e})$ ($\text{supp}(\mathbf{s})$). Edges that are not involved in this set are omitted.}
  \label{fig:trapping}
\end{figure}

\begin{example}\label{ex:select}
Considering the $[[256,32]]$ bicycle code with the given syndrome $\mathbf{s}$, where  $\text{supp}(\mathbf{s})=\{0,2,3,4,5,9,11,15,17,19,20,21,24,25,27,28,30,$
$31,34,37,40,47,50,51,54,58,59,66,69,70,74,78,$
$79,80,82,88,89,90,92,96,97,101,102,103,107,$ $110,111\}$, the corresponding weight-$11$ error pattern $\mathbf{e}$ has the support $\text{supp}(\mathbf{e})=\{14,57,94,101,113,146,150,173,235,238,252\}$.
The estimated support sequence is obtained as $\{140,150,229,6,72,101,88,92,117,149,252,255...\\\}$. With this given syndrome, the iterative decoder induces the trapping set shown in Fig.~\ref{fig:trapping}. Additionally, the black and red margins indicate whether a node belongs to the support or not. Variable nodes $6$ and $255$ each form a cycle-4, with the other two variable nodes $173,252$ and $113,57$, respectively. 
\par Comparing $\text{supp}(\mathbf{e})$ and the estimated support sequence, we observe that $150$ is the first position belonging to $\text{supp}(\mathbf{e})$ that is selected from the estimated support sequence. However, even after selecting position $150$ and reducing the effect of its corresponding syndrome contribution, the decoder still fails to converge. The same situation occurs for error position $101$. The reason is that they will not affect the considered trapping set. For both trials, error position $6$ will always be determined to be in the support of an estimated error pattern during the decoding process. Once position $6$ is selected, position $255$ is subsequently inferred, since all of its neighboring check nodes in the Tanner graph belong to $\text{supp}(\mathbf{s})$. However, neither of them is in $\text{supp}(\mathbf{e})$; the syndrome corresponding to the estimated error pattern must not be consistent with the given syndrome, and the decoder fails. Thus, selecting an error not just from $\text{supp}(\mathbf{e})$ but also involved in the trapping set is crucial in this case. For example, when error position $252$ is selected and reduced, PRE-sRBP can break this trapping set and successfully converge.
\qed
\end{example}

\begin{figure}[!htb]
  \centering
  \includegraphics[width=0.99\columnwidth]{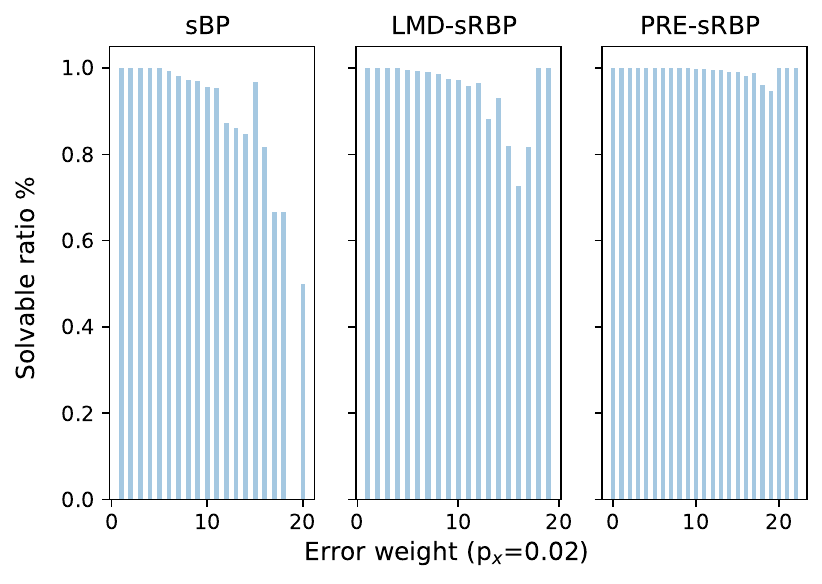}
  \caption{The solvable ratio for different error weights using sBP, LMD-sRBP where $I_{\max}=90$, and PRE-sRBP where $ \lambda_{\max}=15$, $I_t=6$.}
  \label{fig:errWeight400}
\end{figure}

As described in Section \ref{PRE:trapping}, HP codes are more severely impacted by the trapping set. 
We analyze the error weight characteristics of the $[[400,16,6]]$ HP code, initially constructed in \cite{QOSD} and referenced in \cite{Roffe_LDPC_Python_tools_2022}. Owing to its symmetric HP construction, this code contains a large number of quantum trapping sets.
As shown in Fig.~\ref{fig:errWeight400}, LMD-sRBP improves the solvable ratio for certain high-weight error patterns, while for other weights its performance is nearly identical to that of sBP. In contrast, PRE-sRBP consistently achieves solvable ratios exceeding $90\%$ across all evaluated error weights.

\section{Performance Evaluation}\label{EXP}

\subsection{Parameter $\lambda_{\max}$ Set-up for Simulations}

\begin{figure}[ht]
  \centering
  \includegraphics[width=0.99\columnwidth]{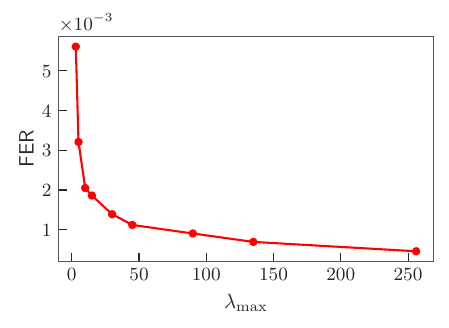}
  \caption{Different $\lambda_{\max}$ values for the [[256,32]] code at $p_x=0.02$ where $I_t=6$.}
  \label{fig:lamdaDecay}
\end{figure}

As illustrated in Example \ref{ex:select}, the selection of $\lambda_{\max}$ plays a crucial role in the performance of PRE-sRBP. In general, a higher value of $\lambda_{\max}$ leads to better performance until the upper bound $N$ is reached, as shown in Fig.~\ref{fig:lamdaDecay}.

However, error patterns that cause decoding failure may incur substantial decoding time when the trials are run sequentially. Considering the latency issue for practical applications, setting a large number for $\lambda_{\max}$ is therefore impractical. Thus, we set an upper bound for $I_{\max}$ for simulations. This configuration introduces a trade-off between $I_t$ and $\lambda_{\max}$, under which an appropriate value of $\lambda_{\max}$ can be determined.
\begin{figure}[ht]
  \centering
  \includegraphics[width=0.99\columnwidth]{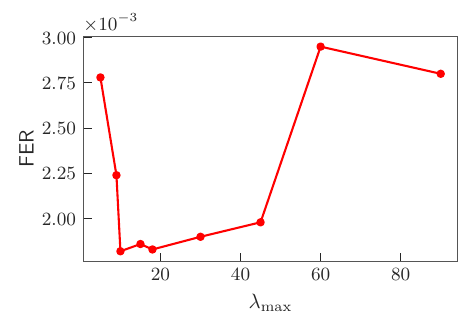}
  \caption{Different $\lambda_{\max}$ selections for the [[256,32]] code at $p_x=0.02$ where $I_{\max}=90$.}
  \label{fig:lamda}
\end{figure}

For instance, considering the $[[256,32]]$ bicycle code at $p_x=0.02$, the frame-error rate (FER) versus the selection of $\lambda_{\max}$ is shown in Fig.~\ref{fig:lamda}. The results indicate that the FER begins to decrease rapidly when $\lambda_{\max}\approx 10$, which is consistent with the observation that the solvable ratio of sRBP-based decoders begins to degrade for error weights larger than 10. However, due to the iteration trade-off, $\lambda_{\max}$ should not exceed 45 to ensure sufficient iterations $I_t$ (at least 2 in this case).

\subsection{Error Rate Simulation}
In each simulation, at least $100$ logical errors are collected, and the maximum number of iterations $I_{\max}=90$ for sBP and syndrome-based IDS decoders unless otherwise specified. In the simulations, sBP uses (\ref{eq_m_c2v}) for C2V updates, and the sLBP processes each check node sequentially.

\begin{figure}[ht]
  \centering
  \includegraphics[width=\columnwidth]{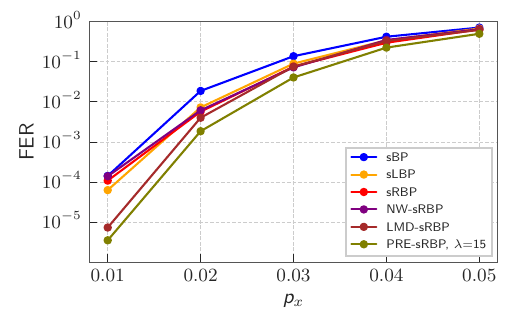}
  \caption{Performance of the $[[256,32]]$ bicycle code using different decoders.}
  \label{fig:FER256}
\end{figure}

Fig.~\ref{fig:FER256} shows the FER performance versus the bit-flip crossover probability $p_x$ using different IDS decoders based on the $[[256,32]]$ bicycle code. The edge pool design significantly impacts performance. For example, NW-sRBP performs worse than the original sRBP and sLBP, contrary to the results for classical LDPC codes. When carefully designed for the edge pool, the performance using LMD-sRBP at $p_x=0.01$ reaches an FER of $7.4\times 10^{-6}$, which improves by almost an order compared to sLBP. By updating diverse edges and proactively reducing the predicted error position, PRE-sRBP where $\lambda_{\max}=15$ achieves the best performance among the considered IDS decoders.

\begin{figure}[ht]
  \centering
  \includegraphics[width=\columnwidth]{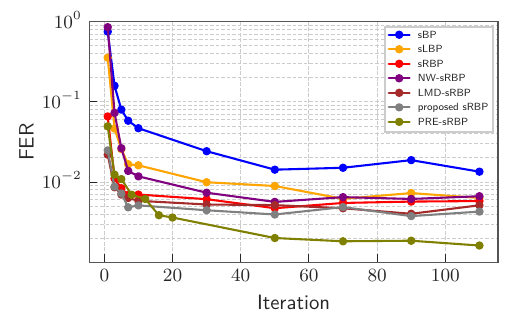}
  \caption{Frame error rate versus $I_{\max}$ on the $[[256,32]]$ bicycle code using different decoders at $p_x=0.02$.}
  \label{fig:ITER256}
\end{figure}

We show the plot for FER versus the maximum number of iterations $I_{\max}$ in Fig.~\ref{fig:ITER256}, which shows that the original sRBP can provide a faster convergence rate than sLBP at $p_x=0.02$. The figure also indicates that the elaborately designed edge pool can substantially improve the convergence behavior. By prioritizing edges neighboring the most recently updated variable nodes, LMD-sRBP achieves a low error rate within only three iterations. Among all decoders considered for this code, the proposed edge pool design (without the PRE mechanism) provides both the fastest convergence rate and the lowest FER, confirming that it effectively explores the code structure while preserving biased updates. For PRE-sRBP, the parameter $\lambda_{\max}$ is adjusted for different $I_{\max}$. The figure shows that PRE-sRBP not only provides a convergence rate on par with the original sRBP in the first few iterations but also the lowest frame-error rate performance after $I_{\max}=16$ where $\lambda_{\max}\geq 5$.

\begin{figure}[ht]
  \centering
  \includegraphics[width=\columnwidth]{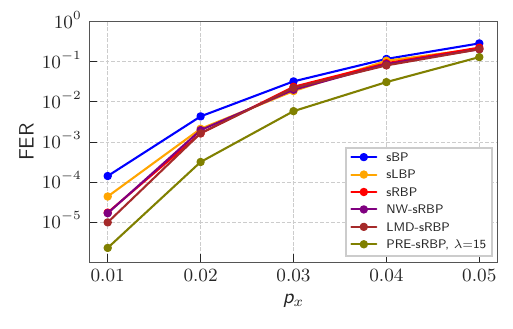}
  \caption{Performance of the $[[126,28,8]]$ generalized bicycle code (\protect\cite[A2]{GBYC}) using different decoders.}
  \label{fig:FER126}
\end{figure}

Fig.~\ref{fig:FER126} shows the performance of the $[[126,28,8]]$ generalized bicycle code using different sRBP-based decoders. The results indicate that using edge-wise decoders can provide a better FER than sBP and sLBP. Whereas, the limitation of further improving performance by designing different edge pools is also reflected in the subtle performance gap between using the original sRBP, NW-sRBP, and LMD-sRBP. Notably, the gap between our proposed decoders and LMD-sRBP for this code is larger than for the $[[256,32]]$ bicycle code, highlighting the significant contribution of the additional PRE mechanism in enhancing decoding performance.

\begin{figure}[ht]
  \centering
  \includegraphics[width=\columnwidth]{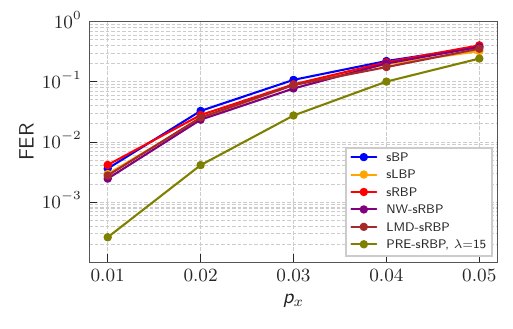}
  \caption{Performance of the $[[400,16,6]]$ HP code using different decoders.}
  \label{fig:FERHGP}
\end{figure}

The structure of the hyper-graph product will lead to multiple quantum trapping sets, which can limit decoding performance \cite{QTRAPPING}. Fig.~\ref{fig:FERHGP} shows that under $I_{\max}=90$, the performance of the $[[400,16,6]]$ HP code using the original sRBP, NW-sRBP, and LMD-sRBP is nearly identical. This observation implies that variations in the edge pool design alone have negligible impact, and a more aggressive mechanism is required for performance improvement. By incorporating the PRE mechanism, PRE-sRBP can yield a significant performance gain compared to other decoders. In Fig.~\ref{fig:FERHGP}, all decoders have converged under $I_{\max}=90$, except for the proposed PRE-sRBP.

\begin{figure}[ht]
  \centering
  \includegraphics[width=\columnwidth]{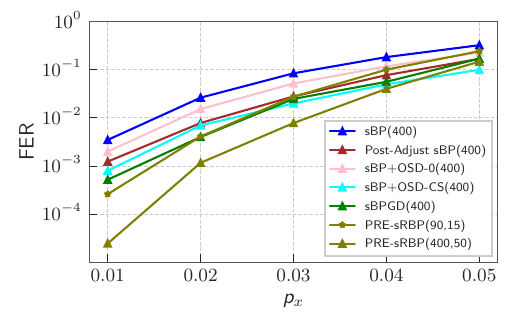}
  \caption{Performance comparison to other sBP-based decoders on the $[[400,16,6]]$ HP code under $I_{\max}=N$.}
  \label{fig:FERHGPITER}
\end{figure}

Under sufficiently large iterations, say $I_{\max}=N$, the error rate using PRE-sRBP can be further improved. In Fig.~\ref{fig:FERHGPITER}, we compare the performance of the $[[400,16,6]]$ HP code using PRE-sRBP to other sBP-based decoders, including the posterior LLR adjustment method (Post-Adjust sBP) \cite{POSTBP}, sBP+OSD \cite{QOSD}, and the sBP-guided decimation (sBPGD) \cite{DECIMATION}. The maximum number of iterations $I_{\max}$ is shown in the brackets, and $\lambda_{\max}$ is shown alongside it for PRE-sRBP. 
Post-Adjust sBP can improve the FER to $10^{-3}$ by flipping the posterior LLR of selected variable nodes at each iteration. sBP+OSD uses the soft output obtained from BP with $I_{\max}=N$ as the reliability auxiliary to search for possible error patterns, where the performance can be improved by increasing the search depth (order). Fig.~\ref{fig:FERHGPITER} shows that sBP+OSD with a zero order (sBP+OSD-0) provides merely a slight performance gain from sBP. A relatively significant increase can be achieved using sBP+OSD with 60 orders by using a combination sweep method (sBP+OSD-CS) at the cost of additional complexity. sBPGD achieves roughly an order of magnitude gain over sBP, requiring $N$ rounds of 400 iterations each.
Remarkably, our proposed PRE-sRBP under $I_{\max}=90$ where $\lambda_{\max}=15$ can outperform those considered decoders when $I_{\max}=400$. Furthermore, the performance can be improved where $I_{\max}=400$, and $\lambda_{\max}=50$, providing about two orders of magnitude improvement compared to sBP.

\begin{figure}[ht]
  \centering
  \includegraphics[width=\columnwidth]{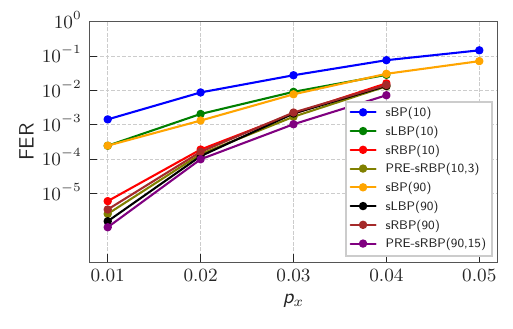}
  \caption{Performance of the $[[144,12,12]]$ bivariate bicycle code using different decoders.}
  \label{fig:BB144}
\end{figure}

Fig. \ref{fig:BB144} shows the performance on $[[144, 12, 12]]$ bivariate bicycle code \cite{IBM} using various sRBP-based algorithms. This code has a distance larger than its stabilizer row weight of 6, making it highly degenerate, where multiple logical errors may correspond to the same syndrome. To determine successful decoding, we use the criterion \(L_z\cdot (\hat{\mathbf{e}} + \mathbf{e}) = 0\), where \(L_z\) is a matrix with each row representing a Z-type logical operator. For this highly degenerate code, the final estimated error pattern is selected as the one with the minimum weight among all convergent trials.

Simulation results indicate that when $I_{\max}=10$, both the original sRBP and PRE-sRBP substantially outperform sBP and sLBP, demonstrating faster convergence. As $I_{\max}$ increases, sLBP surpasses the original sRBP, as described in Section 3.2. In contrast, the proposed PRE-sRBP maintains the fastest convergence while further improving decoding performance at higher $I_{\max}$ by increasing the number of trials, highlighting the benefit of the PRE mechanism for highly degenerate quantum codes.

\subsection{Complexity Anaylsis}
Consider a parity-check matrix whose Tanner graph consists of $M$ check nodes, $N$ variable nodes, and the total number of edges $E=M\cdot\bar{d}_c=N\cdot\bar{d}_v$, where $\bar{d}_c$ and $\bar{d}_v$ denote the average degree of the check nodes and the variable nodes, respectively. We evaluate the complexity in terms of the C2V message update, the V2C message update, the pre-computed C2V update, and the residual comparison per iteration in Table \ref{Tb:complexity}. An iteration for sBP consists of updates to all C2V and V2C edges, i.e., an iteration includes $E$ C2V and V2C updates. For sRBP, an iteration is defined as when the number of C2V updates reaches $E$ due to the unbalanced update distribution specified in Section \ref{IDSintro}. 

In Table \ref{Tb:complexity}, it is shown that both sBP and sLBP have the same number of updates without any pre-computation or comparison. For all sRBP-based decoders, the number of V2C updates and pre-computations are the same since these operations occur in \ref{s:s2}. When a C2V edge is updated, each V2C from $v_{\max}$ (or $v_k$ for NW-sRBP) is then updated except for $c_{\max}$. Thus, there are $\bar{d}_v-1$ V2C updates. Additionally, the new C2V is computed from each check node neighboring $v_{\max}$, except for $c_{\max}$, to its neighboring variable nodes except for $v_{\max}$ (or $v_k$ for NW-sRBP), leading to a total pre-computation of $(\bar{d}_c-1)(\bar{d}_v-1)$.

The difference lies in the residual comparison due to different edge pools $\mathfrak{R}$ in \ref{s:s1} and \ref{s:s3}. For sRBP to determine the C2V update, $\mathfrak{R}$ contains all the check and variable nodes; thus, $E-1$ comparisons are required. For the NW-sRBP, after $E-1$ comparisons, there are $\bar{d}_c$ edges updated from check node $c_{\max}$, i.e., there are $\frac{E-1}{\bar{d}_c}$ comparisons for one C2V update. For one iteration, the comparison is $E\cdot\frac{E-1}{\bar{d}_c}=M(E-1)$. 
In LMD-sRBP, the number of comparisons can be computed in two steps following the method of constructing $\mathfrak{R}_I$ and $\mathfrak{R}$. There are $(\bar{d}_v-1)(\bar{d}_c-1)-1$ comparisons to determine the $v^*_{\max}$ and $\bar{d}_v-1$ to determine the $c^*_{\max}$ neighboring $v^*_{\max}$. Thus, the total number of comparisons for one C2V update is $(\bar{d}_v-1)\bar{d}_c-1$. PRE-sRBP will contain a sequence computation process and the sRBP-based process. In the sRBP-based process, for one C2V update, PRE-sRBP will only compare messages from check nodes neighboring the variable node $v_t$ and those that have not been updated to their neighboring variable nodes excluding $v_j$, resulting in a comparison below $\bar{d}_v(\bar{d}_c-1)-1$.

\subsection{Comparative Evaluation}
In this section, we compare the performance of PER-sRBP with existing sBPGD and sLBP using random order (sLBPRO) \cite{ROLBP} on the $[[882, 24]]$ GHP code \cite{GBYC} under similar complexity. In the following analysis, the complexity of V2C updates is omitted, as it is negligible compared to the C2V update using (\ref{eq_m_c2v}), and each C2V update is assumed to have $O(1)$ complexity.

In the case of sBPGD, the complexity is approximately $O(\mathcal{R}_n\cdot I\cdot E)$, where $\mathcal{R}_n$ is the number of rounds and $I$ represents the number of iterations, respectively. For sLBPRO \cite{ROLBP}, a delicately constructed layered decomposition is used for HP codes. A decomposition is balanced if the number of check nodes in each layer is identical. During each iteration, the decoder processes all layers. For the $[[882, 24]]$ GHP code, a balanced decomposition that each layer includes the indices $i$ and ${i + 3}$, which are both congruent modulo seven, is considered in \cite{ROLBP}. The complexity for sLBPRO is $O(I \cdot 2\cdot E)$, as each check node is updated twice per iteration.

For our PRE-sRBP, the complexity of the PRE mechanism is omitted, as it only operates once prior to decoding. Furthermore, for quantum codes, both the pre-C2V and the residual computation can be disregarded during the first iteration, as only the sign of the C2V update changes while its absolute value remains the same. Additionally, residual comparisons can be efficiently implemented using a priority queue. Taking these optimizations into account, the overall complexity of PRE-sRBP is approximately $O((I-1)\cdot (2+\bar{d}_c\bar{d}_v-\bar{d}_v-\bar{d}_c+\log(\bar{d}_v\bar{d}_c-\bar{d}_v-1))\cdot E)$. 

\begin{figure}[ht]
  \centering
  \includegraphics[width=\columnwidth]{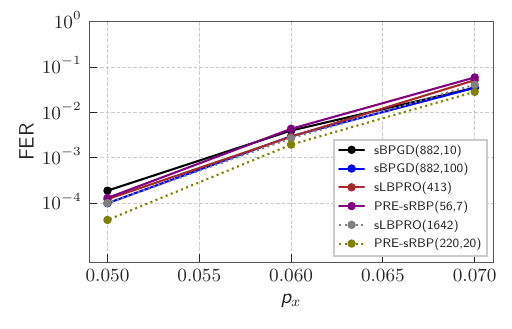}
  \caption{Performance of the $[[882,24]]$ GHP code using different decoders under similar complexity}
  \label{fig:GHP882}
\end{figure}

Fig. \ref{fig:GHP882} shows the performance comparison on the $[[882, 24]]$ GHP code. The first item in the brackets represents the maximum total number of iterations, except for sBPGD, which represents the number of rounds. The second item represents either the maximum number of iterations per round for sBPGD or the number of trials for PRE-sRBP.

The results illustrate that the performance of sBPGD depends on the number of iterations per round; however, both configurations exhibit worst-case complexities exceeding those of the other decoders shown. For sLBPRO, increasing the number of iterations beyond a certain point yields negligible improvement in performance. In contrast, when a sufficient number of candidates and iterations per trial are established, the performance of PRE-sRBP significantly improves compared to sLBPRO under the similar complexity, as shown in the dotted lines.

One appealing aspect of PRE-sRBP for practical applications is its inherent parallelism. Since each trial operates independently, multiple trials can be executed concurrently, and the overall decoding latency is determined solely by the duration of the longest trial. This property is particularly important for quantum error correction, where the decoder must operate quickly enough to prevent data backlog \cite{NISQ}. Current FPGA technology can support latencies on the order of microseconds to nanoseconds \cite{FPGA}, making PRE-sRBP a potential candidate for low-latency hardware implementations.

\begin{table*}[]
\caption{Complexity between sRBP-based decoders per iteration}
\label{Tb:complexity}
\renewcommand\arraystretch{1.2}
\begin{tabularx}{\textwidth}{*{5}{>{\centering\arraybackslash}X}}
\toprule
Schedules  
& C2V update & V2C update & Pre-computation & Residual comparison  \\ 
\midrule
sBP   & $E$      & $E$          & 0   & 0        \\ 
 \hline
sLBP & $E$      & $E$          & 0   & 0          \\ 
 \hline
sRBP       & $E$       & $E(\bar{d}_v-1)$         & $E(\bar{d}_c-1)(\bar{d}_v-1)$  & $E(E-1)$        \\ 
\hline
NW-sRBP        & $E$       & $E(\bar{d}_v-1)$          & $E(\bar{d}_c-1)(\bar{d}_v-1)$   & $M(E-1)$        \\ 
 \hline
LMD-sRBP        & $E$        & $E(\bar{d}_v-1)$          & $E(\bar{d}_c-1)(\bar{d}_v-1)$   & $E[(\bar{d}_v-1)\bar{d}_c-1]$        \\ 
 \hline
PRE-sRBP       & $E$        & $E(\bar{d}_v-1)$          & $E(\bar{d}_c-1)(\bar{d}_v-1)$   & $\leq E[\bar{d}_v(\bar{d}_c-1)-1]$          \\  
\bottomrule
\end{tabularx}
\end{table*}

\section{Conclusions}\label{CONCLU}
In this paper, we have presented several syndrome-based IDS algorithms based on sRBP. Due to the construction constraint and the identical assignment of prior intrinsic information for each variable node, conventional sRBP suffers from the presence of many identical residuals, which limits its decoding performance on QLDPC codes. To tackle this issue, we introduced heuristic modifications that enhance residual diversity and mitigate the impact of quantum trapping sets, leading to faster empirical convergence than sLBP and improved FER performance on the considered bicycle codes. For more challenging code families, including generalized bicycle and HP codes, we showed that stronger prediction-based mechanisms are required to further suppress decoding errors. Motivated by this observation, we proposed the PRE-sRBP decoder, which integrates residual-based scheduling with error prediction and reduction. Although PRE-sRBP does not admit formal convergence guarantees, numerical simulations indicate rapid convergence in practice and demonstrate that it achieves the superior overall error-rate performance among the considered decoders for the examined QLDPC codes, including recent bivariate bicycle constructions.

Future work will concentrate on developing quaternary counterparts of sRBP decoders with improved performance relative to existing quaternary sBP decoders. A key challenge lies in defining an appropriate residual for high-dimensional C2V messages and understanding its influence on the accuracy of edge updates, which will be both intriguing and demanding.

\bibliographystyle{quantum}
\bibliography{IDS_bibliographystyle}

\appendix

\section{Proof of Proposition \ref{prop}} \label{APPENDIX2}
To prove the proposition with a clear description, the sBP in the probability domain, as outlined in \cite{REFBP4}, is used. Denote the message from $v_j$ to $c_i$ and the message from $c_i$ to $v_j$ as $d_{v_j\rightarrow c_i}$ and $\delta_{c_i\rightarrow v_j}$, respectively. The $q_{v_j}^0$ and $q_{v_j}^1$ denote the marginal probability of $v_j$ to be 0 and 1, respectively. The proof is described as follows.

For the sBP in the probability domain, the V2C message $d_{v_j\rightarrow c_i}$ is initialized as $d_{v_j\rightarrow c_i}=(1-p)-p=1-2p$. Then the first iteration that includes three steps:\begin{enumerate}
    \item C2V update: 
    \begin{align*}
    \delta_{c_i\rightarrow v_j}&=(-1)^{s_i}\prod_{v_{j'}}d_{c_i\rightarrow v_{j'}}\\
    &=(-1)^{s_i}(1-2p)^{d_{c_i}-1},
    \end{align*}where $v_{j^{'}}\in \mathcal{N}(c_i)\setminus v_j$. 
    \item V2C update:\\Introduce the intermediary term $r^0_{c_i\rightarrow v_j}=\frac{1+\delta_{c_i\rightarrow v_j}}{2}$ and $r^1_{c_i\rightarrow v_j}=\frac{1-\delta_{c_i\rightarrow v_j}}{2}$, then
    \begin{align*}
    d_{v_j\rightarrow c_i}&=(1-p)\prod_{c_{i'}}r^0_{c_{i'}\rightarrow v_j}-p\prod_{c_{i'}}r^1_{c_{i'}\rightarrow v_j}\\
    &=(1-p)(r^0_{c_i\rightarrow v_j})^{d_{v_j}-1}-p(r^1_{c_i\rightarrow v_j})^{d_{v_j}-1},    
    \end{align*}
    where $c_{i^{'}}\in \mathcal{N}(v_j)\setminus c_i$.
    \item Compute the marginal probability of $v_j$:
    \begin{align*}
    q_{v_j}^0&=(1-p)\prod_{c_i}r^0_{c_i\rightarrow v_j}=(1-p)(r^0_{c_i\rightarrow v_j})^{d_{v_j}} \\
    q_{v_j}^1&=p\prod_{c_i}r^1_{c_i\rightarrow v_j}=p(r^1_{c_i\rightarrow v_j})^{d_{v_j}},
    \end{align*}where $c_i\in \mathcal{N}(v_j)$. 
    \end{enumerate}
    Recall that $\delta_{c_i\rightarrow v_j}$ depends on syndrome position $s_i$, we can further divide $r^0_{c_i\rightarrow v_j}$ and $r^1_{c_i\rightarrow v_j}$ into four terms: $r^0_{c_i\rightarrow v_j,s_i=0},r^0_{c_i\rightarrow v_j,s_i=1},r^1_{c_i\rightarrow v_j,s_i=0}$, and $r^1_{c_i\rightarrow v_j,s_i=1}$. Denote $w^1_{v_j}$ as the cardinality of the set of check nodes neighboring $v_j$ and belonging to the support of the given syndrome, i.e., $w^1_{v_j}=|\{c_i|c_i\in \mathcal{N}(v_j),i\in\text{supp}(\mathbf{s})\}|$, then $q_{v_j}^0$ and $q_{v_j}^1$ can be rewritten as
    \begin{align*}
    q_{v_j}^0&=(1-p)(r^0_{c_i\rightarrow v_j,s_i=1})^{w^1_{v_j}}(r^0_{c_i\rightarrow v_j,s_i=0})^{d_{v_j}-w^1_{v_j}} \\
    q_{v_j}^1&=p(r^1_{c_i\rightarrow v_j,s_i=1})^{w^1_{v_j}}(r^1_{c_i\rightarrow v_j,s_i=0})^{d_{v_j}-w^1_{v_j}}.
    \end{align*}
    By expanding these terms, we can determine 
    \begin{align*}
    r^0_{c_i\rightarrow v_j,s_i=0} &= \frac{1+(1-2p)^{d_c-1}}{2}= r^1_{c_i\rightarrow v_j,s_i=1}  \\
    r^0_{c_i\rightarrow v_j,s_i=1} &= \frac{1-(1-2p)^{d_c-1}}{2}= r^1_{c_i\rightarrow v_j,s_i=0} 
    \end{align*}
    Use $A\triangleq r^0_{c_i\rightarrow v_j,s_i=0}$ and $1-A\triangleq r^1_{c_i\rightarrow v_j,s_i=0}$, then by computing $\frac{q_{v_j}^1}{q_{v_j}^0}$ to decide the hard decision of $v_j$, we can obtain 
    \begin{align*}
    \frac{q_{v_j}^1}{q_{v_j}^0} &= \frac{p}{1-p}(\frac{A}{1-A})^{w^1_j}(\frac{1-A}{A})^{d_{v_j}-w^1_j} \\
    &= \frac{p}{1-p}(\frac{1-A}{A})^{d_{v_j}-2w^1_j}.
    \end{align*}
    Since $A$ is also a function of $p$ given the fixed $d_c$, the hard decision for $v_j$ only depends on $p$ and $d_{v_j}-2w^1_{v_j}$. 
    Then given the fixed $p$, the priority of the estimated $v_j$ belonging to $\text{supp}(\mathbf{e})$ is related to $d_{v_j}-2w^1_{v_j}$.\qed

\end{document}